\documentclass[twocolumn,prb,nopacs,superscriptaddress]{revtex4}
\usepackage{epsf}
\usepackage{amsfonts}
\usepackage[fleqn]{amsmath}
\usepackage{amssymb,yfonts,mathrsfs,bbm}
\usepackage{graphicx}
\usepackage{color}
\renewcommand{\vec}[1]{\mathrm{\mathbf{#1}}}

\begin{document}

\title{Shear and Layer Breathing Modes in Multilayer MoS$_2$}
\author{X. Zhang}
\author{W. P. Han}
\author{J. B. Wu}
\affiliation{State Key Laboratory of Superlattices and Microstructures,
Institute of Semiconductors, Chinese Academy of Sciences, Beijing 100083,
China}
\author{S. Milana}
\affiliation{State Key Laboratory of Superlattices and Microstructures,
Institute of Semiconductors, Chinese Academy of Sciences, Beijing 100083,
China}
\affiliation{Engineering Department, Cambridge University, Cambridge CB3 OFA,UK.}
\author{Y. Lu}
\author{Q. Q. Li}
\affiliation{State Key Laboratory of Superlattices and Microstructures,
Institute of Semiconductors, Chinese Academy of Sciences, Beijing 100083,
China}
\author{A. C. Ferrari}
\email{acf26@eng.cam.ac.uk}
\affiliation{Engineering Department, Cambridge University, Cambridge CB3 OFA,UK.}
\author{P. H. Tan}
\email{phtan@semi.ac.cn}
\affiliation{State Key Laboratory of Superlattices and Microstructures,
Institute of Semiconductors, Chinese Academy of Sciences, Beijing 100083,
China}

\begin{abstract}
We study by Raman scattering the shear and layer breathing modes in multilayer MoS$_2$. These are identified by polarization measurements and symmetry analysis. Their positions change with the number of layers, with different scaling for odd and even layers. A chain model explains the results, with general applicability to any layered material, and allows one to monitor their thickness.

\end{abstract}

\maketitle
The fast progress of graphene research, fuelled by the unique properties of this two dimensional (2d) material, paved the way to experiments on other 2d crystals\cite{novos2d,bonacm,coleman}. There are several layered materials (LMs), studied in the bulk since the sixties\cite{Wilson1969}, retaining their stability down to a single monolayer, and whose properties are complementary to those of graphene. Transition metal oxides\cite{Poizot2000} and metal dichalcogenides have a layered structure\cite{Wilson1969}. Atoms within each layer are held together by covalent bonds, while van der Waals interactions keep the layers together\cite{Wilson1969}. LMs include a large number of systems with interesting  properties\cite{Wilson1969}. E.g., NiTe$_2$ and VSe$_2$ are semi-metals\cite{Wilson1969}, WS$_2$\cite{Prasad1997}, WSe$_2$\cite{Abruna1982}, MoS$_2$\cite{frindt_jap_1966}, MoSe$_2$, MoTe$_2$, TaS$_2$\cite{Clement1978}, RhTe$_2$, PdTe$_2$ are semiconductors\cite{Wilson1969}, h-BN, and HfS$_2$ are insulators, NbS$_2$, NbSe$_2$\cite{Frey2003}, NbTe$_2$, and TaSe$_2$ are superconductors\cite{Wilson1969}, Bi$_2$Se$_3$\cite{Mishra1997}, Bi$_2$Te$_3$\cite{Mishra1997} show thermoelectric properties\cite{Wilson1969} and may be topological insulators\cite{Kane2005}. Similar to graphite and graphene, the LM properties are a function of the number of layers (N). The combinations of such 2d crystals in 3d stacks could offer huge opportunities in designing the functionalities of such heterostructures\cite{novos2d,bonacm}. One could combine conductive, insulating, superconducting and magnetic 2d materials in one stack with atomic precision, fine-tuning the performance of the resulting material\cite{novos2d}, the functionality being embedded in the design of such heterostructures\cite{novos2d}.

Amongst these LMs, MoS$_{2}$ is a subject of intense research because of its electronic\cite{kis_nn_2011} and optical properties\cite{heinz_prl_2010}, such as strong photoluminescence (PL)\cite{heinz_prl_2010,fenwang_nl_2010}, electroluminescence\cite{sundaram}, controllable valley and spin polarization\cite{heinz_nn_2012,Cui12,Feng12}. A single layer MoS$_{2}$ (1L-MoS$_{2}$) consists of two planes of hexagonally arranged S atoms linked to a hexagonal plane of Mo atoms via covalent bonds\cite{heinz_prl_2010,pollmann_prb_2001,wold_prb1987,groot_prb_1987,mattheiss_prb_1973}. In the bulk, individual MoS$_2$ layers are held together by weak van der Waals forces\cite{pollmann_prb_2001,wold_prb1987,groot_prb_1987,mattheiss_prb_1973}. This property has been exploited in lubrication technology\cite{lieber_apl_1991} and, more recently, enabled the isolation of 1L-MoS$_{2}$\cite{kis_nn_2011,heinz_prl_2010,fenwang_nl_2010,novoselov_s_2004}. While bulk MoS$_{2}$ is a semiconductor with a  1.3eV indirect band gap\cite{parkison_jpc_1092}, 1L-MoS$_{2}$ has a 1.8eV direct band gap\cite{heinz_prl_2010,fenwang_nl_2010}. The absence of interlayer coupling of electronic states at the $\Gamma$ point of the Brillouin zone in 1L-MoS$_{2}$\cite{fenwang_nl_2010,galli_jpcc_2007} results in strong absorption and PL bands at$\sim$1.8eV (680nm)\cite{heinz_prl_2010,fenwang_nl_2010}. 1L-MoS$_{2}$ field effect transistors (FETs) show both unipolar\cite{kis_nn_2011} and ambipolar\cite{iwasa_nl_2012} transport, with mobilities$>$500$cm^{2}V^{-1}s^{-1}$ and on-off ratios up to $10^{9}$ [\onlinecite{pdye_IEEE_2012,pdye_acs_2012}]. 1L-MoS$_{2}$ is also a promising candidate for novel optoelectronic devices\cite{sundaram}, such as photodetectors\cite{zhang_acs_2011,orta_oe_2012,im_nl_2012} and light-emitting devices operating in the visible range.
\begin{figure*}
\centerline{\includegraphics[width=180mm,clip]{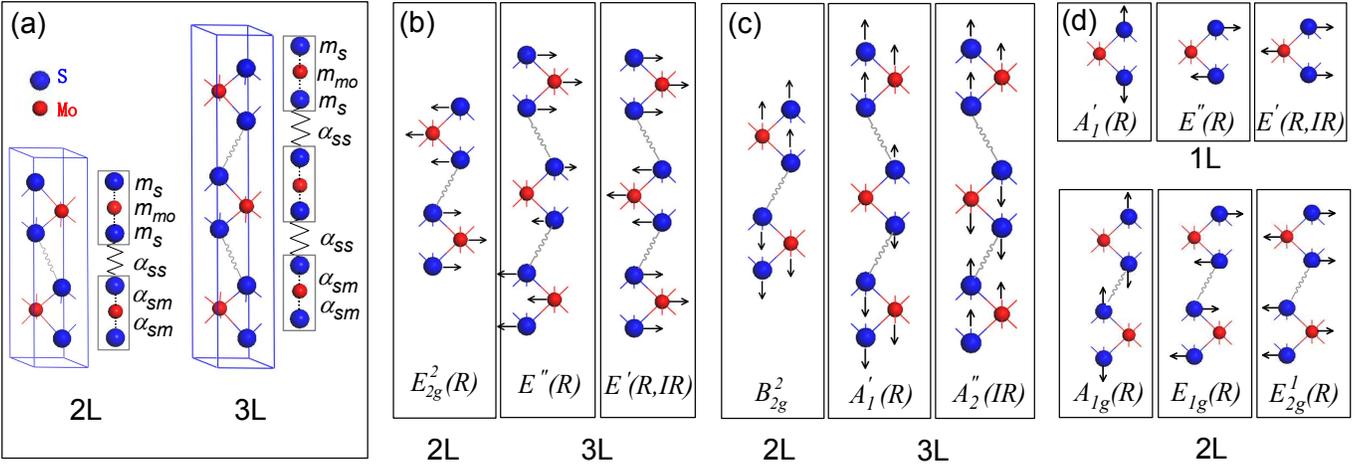}}
\caption{(a) Lattice structure and linear chain models of 2 and 3L-MoS$_2$. Inversion symmetry applies to 2L, not 3L-MoS$_2$. (b) Shear modes and (c) LBMs in 2 and 3L-MoS$_2$. (d) High frequency optical vibration modes for 1 and 2L-MoS$_2$. The symbol under each mode is its irreducible representation. R or IR indicate if the mode is Raman, or Infrared active, or both.}
\label{fig:1}
\end{figure*}

Raman spectroscopy is the prime non-destructive characterization tool for carbon materials\cite{ferraribook,ferrari07}, in particular graphite\cite{Tuinstra1970,Nemanich1979,wangyan1990,ferrari07}, single\cite{ferrari07,ferrari06} and multilayer\cite{ferrari07,ferrari06} graphene. The Raman spectrum of graphene consists of two fundamentally different sets of peaks. Those, such as D, G, 2D, etc, due to in-plane vibrations\cite{ferraribook,Tuinstra1970,ferrari07}, and others, such as the shear (C) modes\cite{tan12} and the layer breathing (LB) modes (LBMs)\cite{tony nl 2012,tony nl 2012b,sato2011}, due to relative motions of the planes themselves, either perpendicular or parallel to their normal. Albeit being an in-plane mode, the 2D peak is sensitive to N since the resonant Raman mechanism that gives rise to it is closely linked to the details of the electronic band structure\cite{ferrari06,ferrari07,Latil}, the latter changing with N\cite{Koshino,McCann}, and the layers relative orientation\cite{Latil}. On the other hand, the C modes and LBMs are a direct probe of N\cite{tan12,tony nl 2012,tony nl 2012b}, since the vibrations themselves are out of plane, thus directly sensitive to N. The success of Raman scattering in characterizing graphene prompted the community to extend this technique to other LMs, from bulk to monolayer\cite{tan12,Heinz10b,sekine1980,reich05,jzhang2011,Plechinger2012,cuixd12}. E.g., the Raman spectrum of bulk MoS$_2$ consists of two main peaks$\sim$382, 407cm$^{-1}$[\onlinecite{wiet70,wiet71}]. These are assigned to $E_{2g}^1$ (in-plane vibration) and $A_{1g}$ (out of plane vibration) modes\cite{wiet70,wiet71}. The $E_{2g}^1$ mode red-shifts, while the $A_{1g}$ mode blue shifts with increasing N\cite{Heinz10b,Li12}. The $E_{2g}^1$ and $A_{1g}$ modes have opposite trends when going from bulk MoS$_2$ to 1L-MoS$_{2}$, so that their difference can be used to monitor N\cite{Heinz10b}. The $E_{2g}^1$ shift with N may be attributed to stacking-induced structure changes or long-range Coulombic interlayer interactions\cite{Heinz10b,Li12}, while the A$_{1g}$ shift is due to increasing restoring forces as additional layers are added\cite{Heinz10b,Li12}, however, further work is still needed to fully clarify and assign these trends, but this is not the subject of the present investigation.

Instead, our focus here is on the C and LB modes that appear in the low frequency region in the various LMs\cite{c44}. These have been extensively studied in multilayer graphene\cite{tan12,tony nl 2012,tony nl 2012b,Michel12}. Unlike graphite and graphene, most LMs consist of more than one atomic element. E.g., each MoS$_2$ layer contains one Mo plane sandwiched by two S planes, while Bi$_2$Se$_3$ contains two Bi and three Se planes. This makes their lattice dynamics more complex than multilayer graphene, starting from the symmetry and force constants. Even NL-MoS$_2$ belong to point group D$_{6h}$ with inversion symmetry, while odd NL-MoS$_2$ correspond to D$_{3h}$ without inversion symmetry\cite{pointdif}. There are a few reports on C and LBMs in LMs other than graphene. Ref.[\onlinecite{richard74}] reported them in bulk samples as:$\sim$21.5cm$^{-1}$ (C),32.5cm$^{-1}$(C), $\sim$50cm$^{-1}$(LBM) for As$_2$Se$_3$ at 15K; $\sim$27cm$^{-1}$(C),38cm$^{-1}$(C), $\sim$60cm$^{-1}$(LBM) for As$_2$S$_3$ at 15K; $\sim$34cm$^{-1}$ (C), $\sim$56cm$^{-1}$ (LBM) for MoS$_2$; $\sim$22cm$^{-1}$ (C) for GaS; $\sim$56cm$^{-1}$ (C) for GaSe. Only Refs[\onlinecite{Plechinger2012,cuixd12}] reported some of these for non-bulk samples. In particular, Ref.[\onlinecite{Plechinger2012}] studied the Raman spectrum of one shear mode for 2, 3, 5, 6 and 10L-MoS$_2$. Ref.[\onlinecite{cuixd12}] observed only one set of C and LB modes for 2 to 6L-MoS$_2$ and 9L-MoS$_2$. Both did not consider the symmetry difference between odd- and even-NL MoS$_2$, e.g. they assigned the C mode in odd-NL MoS$_2$ as $E_{2g}^2$, but, as we show later, this is instead $E^{'}$. Ref.[\onlinecite{Plechinger2012}] suggested that the scaling rule of the C mode in multilayer graphene\cite{tan12} cannot be extended to few layer MoS$_2$, opposite to the results presented in Ref.[\onlinecite{cuixd12}]. Ref.[\onlinecite{cuixd12}] wrote that LBMs scale as 1/N, as predicted by Ref.[\onlinecite{Luo96}] with an assumption of strong coupling between layers and substrate. However, it is not clear whether such strong coupling actually exists. Furthermore, even though LBMs are optical modes, an acoustic atomic displacement for such modes was presented in Fig.1b of Ref.[\onlinecite{cuixd12}], with no symmetry analysis. Therefore, all symmetries, force constants, possible role of interactions between layers and substrate, and mode scaling with N still need to be fully understood. More importantly, it would be desirable to establish a general model to describe the evolution of C and LB modes with N in any LMs, not just MoS$_2$.

Here we measure the shear and layer breathing modes for NL-MoS$_2$ up to 19L-MoS$_2$, and bulk MoS$_2$. We identify several groups of modes with frequencies dependent on N. Samples with even and odd N show different scaling laws with N, due to their different symmetry. A simple chain model can account for the observed trends, and can be extended to other LMs.
\begin{figure*}
\centerline{\includegraphics[width=180mm,clip]{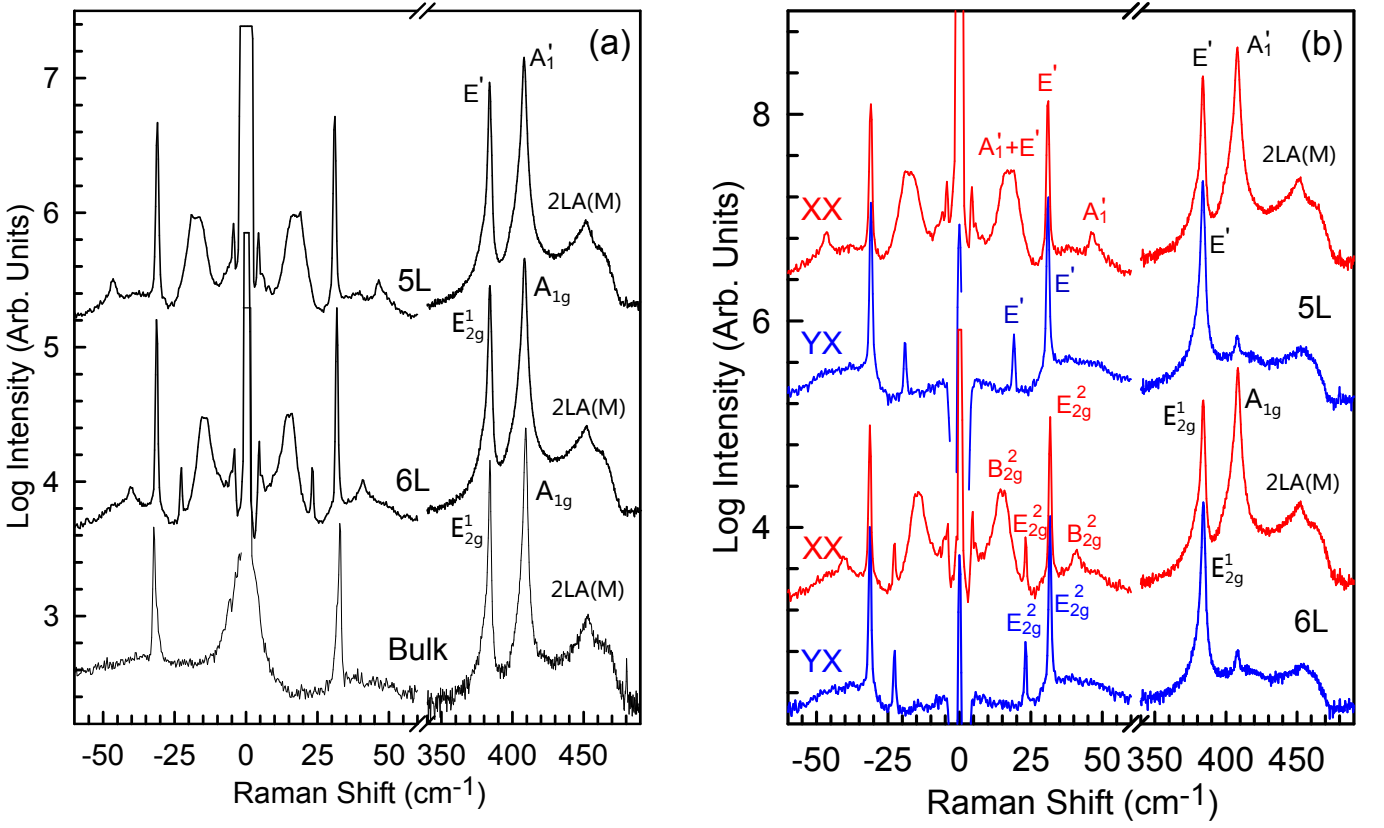}}
\caption{(a) Raman spectra of 5, 6L and bulk MoS$_2$. (b) Raman spectra of 5 and 6L-MoS$_2$ measured for XX (red) and YX (blue) polarizations. The irreducible representation of each mode is indicated.}
\label{fig:2}
\end{figure*}

NL-MoS$_2$ samples are produced from bulk MoS$_2$ (SPI Supplies) by mechanical exfoliation, following a similar procedure to that used for graphene layers\cite{Novoselov2005,bonacm}. NL-MoS$_2$ are supported on a Si wafer with 93nm SiO$_2$, which is used as substrate in order to make the samples visible by eye. The layer thickness is determined by optical contrast\cite{Casiraghi07} and atomic force microscopy\cite{Li12}. Raman measurements are performed using a Jobin-Yvon HR800 system equipped with a liquid nitrogen cooled charge-coupled detector. The excitation wavelength is 532nm from a diode-pumped solid-state laser. A power$\sim$0.23mW is used to avoid sample heating. The laser plasma lines are removed using a BragGrate bandpass filter (OptiGrate Corp), as these would appear in the same spectral range as the modes of interest. The Rayleigh line is suppressed using four BragGrate notch filters with an optical density 3 and a spectral bandwidth$\sim$5-10cm$^{-1}$. This configuration is similar to that used in Ref.[\onlinecite{tan12}] for multi-layer graphene. The spectral resolution is$\sim$0.6cm$^{-1}$, as estimated from the the Rayleigh peak full-width at half-maximum (FWHM).

Bulk MoS$_2$ and 2L-MoS$_2$ belong to the space group P6$_3$/$_{mmc}$ (point group D$_{6h}$)\cite{coordination}, with unit cell consisting of two Mo atoms in sites with point group D$_{3h}$, and four S atoms in sites with point group C$_{3v}$\cite{wiet70}, as shown in Fig.\ref{fig:1}a. There are 18 normal vibration modes\cite{coordination}. The factor group of bulk and 2L-MoS$_2$ at $\vec{\Gamma}$ is D$_{6h}$, the same as the point group\cite{zhangguangyin1991}. The atoms site groups are a subgroup of the crystal factor group\cite{zhangguangyin1991}. The correlation\cite{wiet70} of the Mo site group D$_{3h}$, S site group C$_{3v}$, and factor group D$_{6h}$ allows one to derive the following irreducible representations for the 18 normal vibration modes at $\vec{\Gamma}$\cite{wiet70,ataca11}: $\vec{\Gamma}$= $A_{1g}+2A_{2u}+2B_{2g}+B_{1u}+E_{1g}+2E_{1u}+2E_{2g}+E_{2u}$, where $A_{2u}$ and $E_{1u}$ are translational acoustic modes, $A_{1g}$, $E_{1g}$ and $E_{2g}$ are Raman active, $A_{2u}$ and $E_{1u}$ are infrared (IR) active. The $E_{1g}$ and $A_{1g}$ modes and one of the doubly degenerate E$_{2g}$ modes, E$_{2g}^1$, as shown in Fig.\ref{fig:1}d for 2L and bulk MoS$_2$, give rise to Raman modes above 200cm$^{-1}$[\onlinecite{Heinz10b}]. Only $A_{1g}$ ($\sim$408cm$^{-1}$ in bulk and $\sim$405cm$^{-1}$ in 2L-MoS$_2$) and E$_{2g}^1$ ($\sim$382cm$^{-1}$ in bulk and $\sim$383cm$^{-1}$ in 2L-MoS$_2$) can be observed when the laser excitation is normal to the sample basal plane\cite{Heinz10b}. The other doubly degenerate E$_{2g}$ mode, E$_{2g}^2$, and one B$_{2g}$ mode, B$_{2g}^2$, are shear and LB modes\cite{wiet70,wiet71,Davydov}. $E_{2g}^2$ corresponds to a rigid-layer displacement perpendicular to the c axis (C modes), while B$_{2g}^2$ corresponds to rigid-layer displacements parallel to the c axis (LBMs), as shown in Figs\ref{fig:2}b,c for 2L and bulk MoS$_2$.

1L-MoS$_2$ has $D_{3h}$ symmetry, with three atoms per unit cell\cite{coordination}. The irreducible representation of $D_{3h}$\cite{coordination,ataca11} gives: $\vec{\Gamma}$= $2A_2^{''}$+$A_1^{'}$+$2E^{'}$+$E^{''}$, with $A_2$$^{''}$ and $E^{'}$ acoustic modes, $A_2^{''}$ IR active, $A_1^{'}$ and $E^{''}$ Raman active, and the other $E^{'}$ both Raman and IR active. The $E^{'}$ and $A_1^{'}$ modes, Fig.\ref{fig:1}d, were previously detected in the Raman spectra of 1L-MoS$_2$ at$\sim$384 and$\sim$403cm$^{-1}$[\onlinecite{Heinz10b}]. Of course, no rigid-layer vibrations can exist in 1L-MoS$_2$.
\begin{figure*}
\centerline{\includegraphics[width=180mm,clip]{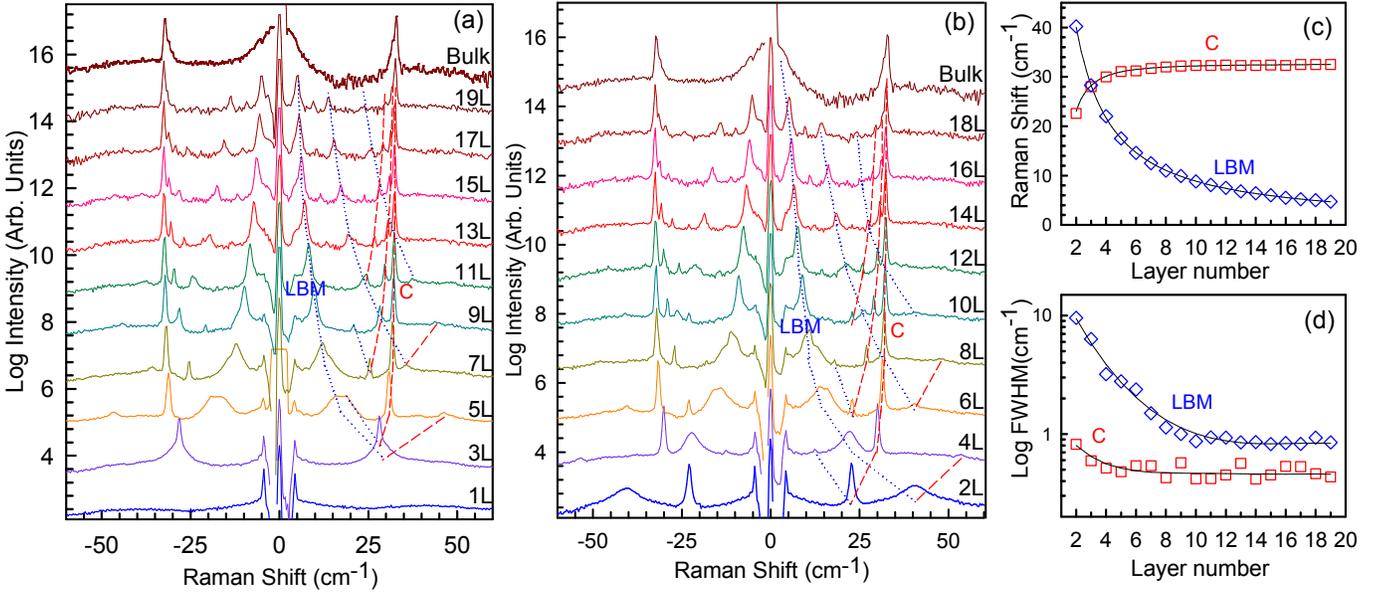}}
\caption{(a) Stokes and anti-Stokes Raman spectra of ONL-MoS$_2$ in the low frequency range. (b) Stokes and anti-Stokes Raman spectra of ENL-MoS$_2$. The spectrum of bulk MoS$_2$ is also included in (a,b). Dashed and dotted lines in (a,b) are guides to the eye. (c) Position of typical C and LB modes as a function of N. (d) FWHM of C and LBM as a function of N. Solid lines in (c,d) are guides to the eye}
\label{fig:3}
\end{figure*}

3L-MoS$_2$ has the same point group ($D_{3h}$) as 1L-MoS$_2$, with $A_1^{'}$ and $A_2^{''}$ corresponding to LBMs, and $E^{'}$ and $E^{''}$ being shear modes, Figs\ref{fig:1}b,c. Systems with even N belong to point group D$_{6h}$ (with inversion symmetry), while odd N correspond to D$_{3h}$ (without inversion symmetry)\cite{pointdif}. Therefore, it is convenient to denote each mode of NL-MoS$_2$ by the corresponding irreducible representation according to the their point group, and then determine if they are Raman, IR active, or inactive.

NL-MoS$_2$ has 9N-3 optical modes: 3N-1 are vibrations along the c axis, and 3N-1 are doubly degenerate in-plane vibrations. For rigid-layer vibrations, there are N-1 LBMs along the c axis, and N-1 doubly degenerate shear modes perpendicular to it. When N is even, there are 0 Raman active LBMs and $\frac{N}{2}$ doubly degenerate shear modes. When N is odd, there are $\frac{N-1}{2}$ LBMs and N-1 doubly degenerate shear modes. The inter-layer distance in LMs is much larger than the in-plane bond length, e.g. in MoS$_2$ the inter-layer distance is$\sim$6.7\AA, while the in-plane bond length is$\sim$3.2\AA\cite{windom}. Thus, the in-plane optical modes may not strongly depend on N. However, the interlayer coupling dominates the lattice dynamics of the rigid-layer vibrations, so that LB and shear modes will be very sensitive to N. E.g., Fig.\ref{fig:2}a shows the Raman spectra of 5L, 6L and bulk MoS$_2$. In the high frequency region above 200cm$^{-1}$, the $E_{2g}^1$ ($\sim$384cm$^{-1}$), $A_{1g}$ ($\sim$409cm$^{-1}$), 2LA(M) ($\sim$453cm$^{-1}$) and $A_{2u}$ ($\sim$463cm$^{-1}$) modes are detected both in bulk and 6L-MoS$_2$. Although the notation in 5L-MoS$_2$ is different from 6L and bulk MoS$_2$ because of the crystal symmetry, the modes ($E^{'}$, $A_1^{'}$) are also observed in 5L-MoS$_2$. The lineshape and peak positions in the high frequency region for 5L and 6L-MoS$_2$ is almost identical, both being similar to bulk MoS$_2$. In the low frequency region below 100cm$^{-1}$, there is only one Raman peak$\sim$33cm$^{-1}$, i.e., $E_{2g}^2$, in bulk MoS$_2$. However, as discussed above, there should exist 6 Raman active modes for 5L and 3 for 6L-MoS$_2$. Of these, 4 and 3 shear modes should be doubly degenerate for 5L and 6L-MoS$_2$, respectively. Experimentally, we observe 3 modes below 60 cm$^{-1}$ in 5L and 4 in 6L-MoS$_2$, as shown in Fig.\ref{fig:2}a.

The LB ($A_1^{'}$) and shear ($E^{'}$, $E^{''}$) Raman tensors in odd NL-MoS$_2$ (ONL-MoS$_2$) and shear ($E_{2g}^2$) Raman tensor in even NL MoS$_2$ (ENL-MoS$_2$) are\cite{tensor,polar11}:
\[A_1^{'} (LB,ONL):
\left [ \begin{array}{clcr}
 a & 0 & 0 \\
 0 & a & 0 \\
 0 & 0 & b
\end{array} \right ],
\]
\[ E^{'} (shear,ONL):
\left [ \begin{array}{clcr}
 0 & d  & 0 \\
 d & 0 & 0 \\
 0 & 0  & 0
\end{array} \right ],
\left [ \begin{array}{clcr}
 d  & 0  & 0 \\
 0  & -d & 0 \\
 0  & 0  & 0
\end{array} \right ],
\]
\[ E^{''} (shear,ONL):
\left [ \begin{array}{clcr}
 0 & 0 & 0 \\
 0 & 0 & c \\
 0 & c & 0
\end{array} \right ],
\left [ \begin{array}{clcr}
 0 & 0 & -c \\
 0 & 0 & 0 \\
-c & 0 & 0
\end{array} \right ];
\]
\[ E_{2g}^2 (shear,ENL):
\left [ \begin{array}{clcr}
 0 & d & 0 \\
 d & 0  & 0 \\
 0 & 0  & 0
\end{array} \right ],
\left [ \begin{array}{clcr}
 d & 0 & 0 \\
 0 & -d  & 0 \\
 0 & 0  & 0
\end{array} \right ].
\]
\noindent
We do not discuss the LBM ($B_{2g}^2$) in ENL-MoS$_2$ since it is Raman inactive. These tensors show that, in backscattering, the $A_1^{'}$ modes in 5L-MoS$_2$ should appear only under unpolarized XX configuration, and $E^{'} $ should exist under both unpolarized XX and polarized YX, while $E^{''}$ should not appear for either XX or YX.  Here XY indicates two mutually perpendicular axes within the basal plane of NL-MoS$_2$, the first being the polarization direction of the incident laser, the second the analyzer's polarization. For 6L-MoS$_2$ under back-scattering, the $E_{2g}^2$ modes exist for both XX and YX configurations.
\begin{figure*}
\centerline{\includegraphics[width=180mm,clip]{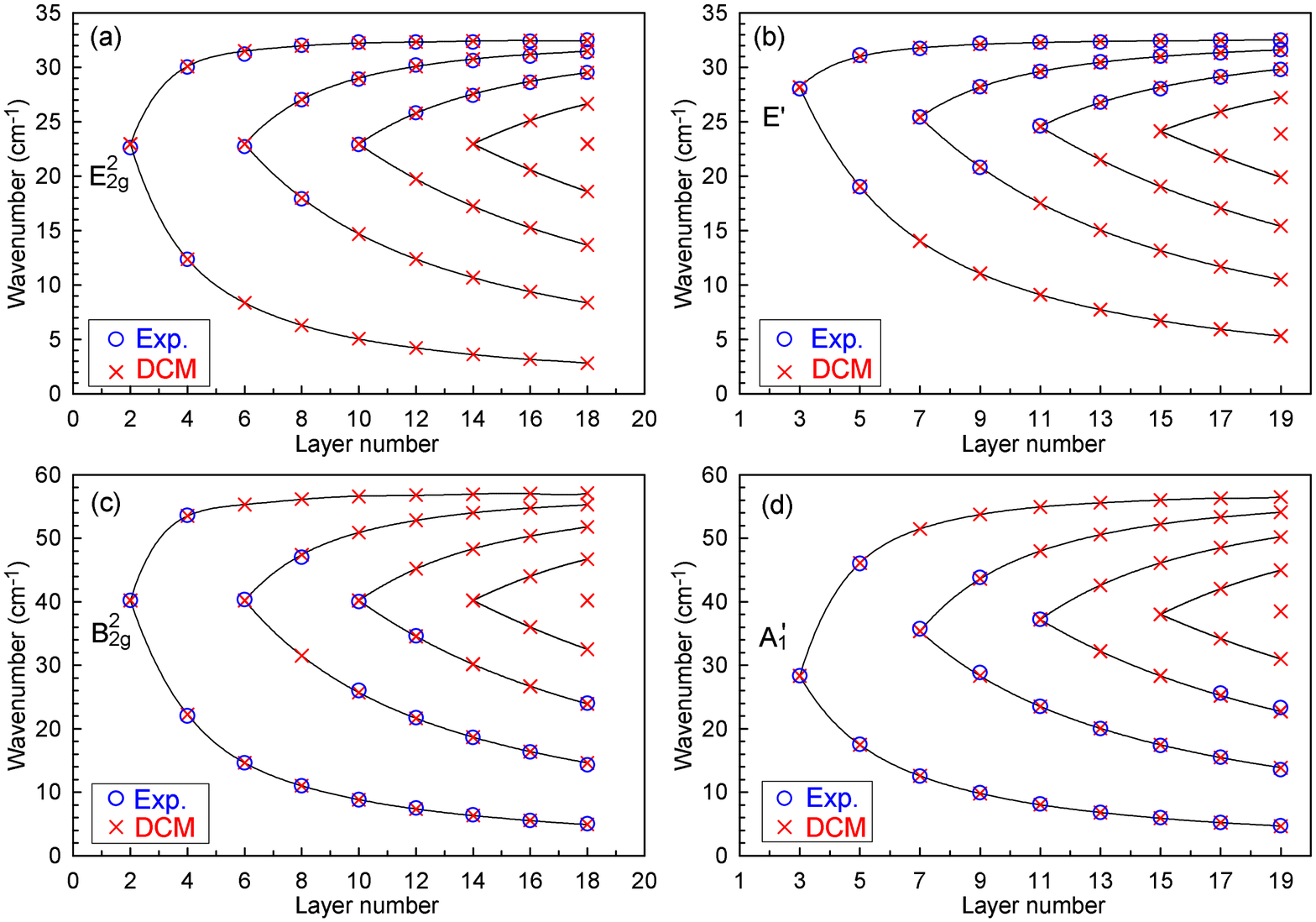}}
\caption{Position of shear modes as a function of N for (a) ENL- and (b) ONL-MoS$_2$.  Position of LBMs as a function of N for (c) ENL- and (d) ONL-MoS$_2$. The blue open circles are the experiment data. The red crosses are the diatomic chain model calculations. The irreducible representation of each mode is also indicated. Solid lines are guides to the eye.}
\label{fig:4}
\end{figure*}

In Fig.\ref{fig:2}b two sharp peaks are observed under both XX and YX configurations at$\sim$19 and$\sim$30cm$^{-1}$ for 5L, and$\sim$23 and$\sim$32cm$^{-1}$ for 6L-MoS$_2$. According to the symmetry analysis discussed above, we assign these to $E^{'}$ in 5L and $E_{2g}^2$ in 6L-MoS$_2$. Two broad peaks are observed for XX measurements at$\sim$17 and 47cm$^{-1}$ for 5L-MoS$_2$, which we assign to $A_1^{'}$. Note that the lower $E^{'}$ mode of 5L-MoS$_2$ at$\sim$19cm$^{-1}$ cannot be fully resolved for XX measurements due to the presence of the broad $A_1^{'}$ mode$\sim$17cm$^{-1}$. We also detect two Raman modes$\sim$15 and 41cm$^{-1}$ in 6L-MoS$_2$ consistent with what should be optically silent $B_{2g}^2$ LBMs, as discussed later. The symmetry, polarization and position are summarized in Table I for the shear and LB modes of 5 and 6L-MoS$_2$. Although the in-plane modes in 5 and 6L-MoS$_2$ above 200cm$^{-1}$ are almost identical in frequency and lineshape, the shear and LB mode positions below 100cm$^{-1}$ are different. The frequencies of all 5L-MoS$_2$ LBMs are higher than in 6L-MoS$_2$, while all shear modes are lower.
\begin{table}
\caption{Symmetry, polarization and experimental (exp.) positions of shear and LB modes in 5 and 6L-MoS$_2$}
\begin{center}
\begin{tabular}{c|c|c|c|c|c}
  \hline\hline
  \multicolumn{2}{c|}{} &\multicolumn{2}{c|}{Shear mode} &\multicolumn{2}{c}{LBM} \\ \hline
   & mode &~E$'$(R,IR)~&~E$'$$'$(R)~&~A$_1'$(R)~&~A$_2'$(IR)~\\ \cline{2-6}
   ~~5L~~ & polarization &XX,YX &XZ,YZ& XX& - \\ \cline{2-6}
    & exp.(cm$^{-1}$) &19,30 &- & 17,47 & - \\ \hline
    & mode &\multicolumn{2}{c|}{E$_{2g}^2$(R)} & \multicolumn{2}{c}{B$_{2g}^2$(silent)} \\ \cline{2-6}
  ~~~6L~~~ & polarization &\multicolumn{2}{c|}{XX,YX} & \multicolumn{2}{c}{-} \\ \cline{2-6}
   & exp.(cm$^{-1}$) &\multicolumn{2}{c|}{23,32} & \multicolumn{2}{c}{15,41(XX)} \\
  \hline\hline
\end{tabular}
\end{center}
\end{table}

Figs\ref{fig:3}a,b show the low frequency Raman measurements for NL-MoS$_2$, with N=1...19, as well as bulk MoS$_2$. Since the point group of ONL-MoS$_2$ ($D_{3h}$) is different from ENL-MoS$_2$ ($D_{6h}$), we plot the Raman spectra of ONL- (Fig.\ref{fig:3}a) and ENL- (Fig.\ref{fig:3}b) MoS$_2$ in two panels. Bulk MoS$_2$ is included both in ONL-MoS$_2$ and ENL-MoS$_2$ panels because we cannot distinguish its parity in a (non-infinite) bulk sample. Of course there are no LB nor shear modes in 1L-MoS$_2$, as confirmed in Fig.\ref{fig:3}a. The two spikes$\sim$4.55cm$^{-1}$, with weaker intensity for thicker MoS$_2$ flakes, are due to Brillouin scattering of the LA mode from the Si substrate\cite{kuok2000}. This is confirmed by determining the elastic constant c$_{11}$ from c$_{11}$=$\rho$$\nu^2$$\pi^2$/($\eta^2$+$\kappa^2$)$k_0^2$, where $\rho$ is the Si density, $\nu$ is the LA mode frequency, k$_0$=2$\pi$/$\lambda_0$, $\lambda_0$ is the incident light wavelength and ($\eta$+i$\kappa$) is the Si complex refractive index. The c$_{11}$ determined from our Raman measurements ($\sim$1.65$\times$10$^{11}$Pa) is consistent with 1.66$\times$10$^{11}$Pa measured by ultrasonic wave propagation\cite{c11}. In 2 and 3L-MoS$_2$, two Raman peaks are observed, with two peaks overlapping in 3L-MoS$_2$. More Raman peaks are observed for thicker MoS$_2$ flakes.

We classify all low frequency Raman peaks into two categories. Those that stiffen for increasing N, linked by dashed lines, and those softening with N, linked by dotted lines. One shear mode stiffens from$\sim$22.6cm$^{-1}$ in 2L to 32.5cm$^{-1}$ in 19L-MoS$_2$, while one LBM softens from$\sim$40.1cm$^{-1}$ in 2L to 4.7cm$^{-1}$ in 19L-MoS$_2$, Fig.\ref{fig:3}c. The two sets of modes have different FWHM, Fig.\ref{fig:3}d. In 2L-MoS$_2$, FWHM(LBM$\sim$40.1cm$^{-1}$) is$\sim$9.6cm$^{-1}$, much larger than that ($\sim$0.8cm$^{-1}$) of the shear mode at 22.6cm$^{-1}$. All data for ENL- and ONL-MoS$_2$ are summarized in Figs\ref{fig:4}a-d. According to group analysis, there should be no Raman active LBMs in ENL-MoS$_2$. However, a set of peaks are observed in ENL-MoS$_2$, with the same polarization behavior as LBMs in ONL-MoS$_2$, e.g. the two Raman modes$\sim$15 and 41cm$^{-1}$ in 6L-MoS$_2$ in Fig.\ref{fig:2}b. Because their measured positions match well those predicted for LBMs, i.e. the $B_{2g}^2$ modes in ENLs as discussed later, they are included in Fig.\ref{fig:4}c.

Since a MoS$_2$ layer consists of two types of atoms, S and Mo, we implement a diatomic chain model (DCM) to explain the data. Fig.\ref{fig:1}a shows the ball and stick model for 2 and 3L-MoS$_2$. Only two force constants are needed to describe the vibrations: $\alpha_{ss}$ and $\alpha_{sm}$, where $\alpha_{ss}$ is the force constant per unit area between two nearest S planes in two adjacent layers, and $\alpha_{sm}$ that between the nearest S and Mo planes within a MoS$_2$ layer. Their components perpendicular to the basal plane, $\alpha_{ss}^\perp$ and $\alpha_{sm}^\perp$, determine the LBMs lattice dynamics, while those parallel to the basal plane, $\alpha_{ss}^\parallel$ and $\alpha_{sm}^\parallel$, determine the shear modes dynamics. The reduced mass for a S (Mo) plane, $m_{S}$ ($m_{Mo}$), is its atomic mass per unit area. In MoS$_2$, $m_{S}=0.6\times {1}0^{-7}$g/cm$^2$ and $m_{Mo}=1.8\times {1}0^{-7}$g/cm$^2$. $\alpha_{sm}^\perp$ and $\alpha_{sm}^\parallel$ can be estimated from the high frequency $A_1^{'}$ and $E^{'}$ modes of 1L-MoS$_2$, for which 9$\times$9 dynamical matrices can be constructed and solved analytically. We get $\omega_{A_1^{'}}=(1/2\pi c)\sqrt{2\alpha_{sm}^\perp/\mu}$, with $\mu=2m_S$. The atom displacement eigenvectors show that the vibration directions of the two external S atoms are opposite along the c axis, while the center Mo atom stays still, as shown in Fig.\ref{fig:1}d for lL-MoS$_2$, corresponding to a spring connected by two S atoms with a force constant per unit area $2\alpha_{sm}^\perp$, since the Mo atom stays still at the spring equilibrium position. We measure $\omega_{A_1^{'}}\sim$403cm$^{-1}$ in 1L-MoS$_2$. This gives $\alpha_{sm}^\perp=3.46\times10^{21}N/m^3$. We also get $\omega_{E^{'}}=(1/2\pi c) \sqrt{2\alpha_{sm}^\parallel/\mu}$, where $1/\mu=1/m_{M_0}+1/(2m_S)$. The atom displacement eigenvectors indicate that the vibration directions of the two S atoms are opposite to the center Mo atom, along the basal plane, as shown in Fig.\ref{fig:1}d for 1L-MoS$_2$, corresponding to a spring connected by two S atoms and one Mo atom with a force constant per unit area $2\alpha_{sm}^\parallel$. From the experimental 384cm$^{-1}$, we get $\alpha_{sm}^\parallel=1.88\times10^{21}N/m^3$. Note that the weak interaction of the two S planes in the MoS$_2$ layers is not included because the S-S plane distance is twice the S-Mo one.

To understand the shear modes of NL-MoS$_2$, the layer coupling between two nearest S planes in the two adjacent layers should be included. 3N$\times$3N dynamical matrices can be constructed for NL-MoS$_2$. By numerically solving the eigen-equation for NL-MoS$_2$, we get the eigenfrequencies and corresponding eigenvectors. By fitting these to our experimental data we get: $\alpha_{ss}^\perp$=$8.90\times10^{19}N/m^3$ and $\alpha_{ss}^\parallel$=$2.82\times10^{19}N/m^3$. Multiplying $\alpha_{ss}^\perp$ and $\alpha_{ss}^\parallel$ by the unit cell area gives the interlayer force constants, $k_{ss}^{shear}$=2.5N/m and $k_{ss}^{LBM}$=7.8N/m. They agree well with those for bulk samples reported in Ref.[\onlinecite{richard74}] ($k_{1}^{shear}$=2.7N/m and $k_{1}^{comp}$=7.4N/m), derived by considering S-Mo-S as a rigid-layer mass unit, and deducing the force constants from $\omega=\sqrt{k_1/\mu_1}$, with $\omega$ the rigid-layer frequency and $\mu_1$ the reduced mass. Multiplying $\alpha_{ss}^\parallel$ by the equilibrium distance between two adjacent MoS$_2$ layers gives a shear modulus$\sim$18.9GPa, in good agreement with that measured for bulk MoS$_2$\cite{Feldman81}, from phonon dispersion curves determined by neutron scattering, and X-ray measurements of the linear compressibilities.

The eigenvectors of the rigid-layer vibrations in 2 and 3L-MoS$_2$ derived from the corresponding eigen-equations are depicted in Figs\ref{fig:1}b,c. Applying symmetry analysis to the corresponding NL-MoS$_2$ eigenvectors, we assign the irreducible representations of the corresponding point group to each mode. The eigenfrequencies of Raman active rigid-layer vibrations for $E_{2g}^2$ (C modes in ENL-MoS$_2$), $E^{'}$ (C modes in ONL-MoS$_2$), and $A_1$$^{'}$ (LBMs in ONL-MoS$_2$) are summarized in Figs.\ref{fig:4}a,b,d, respectively. The eigenfrequencies of the Raman inactive $B_{2g}^2$ (LBMs) in ENL-MoS$_2$ are also included in Fig.\ref{fig:4}c. As illustrated in Fig.\ref{fig:4}, the model calculations are in good agreement with experiments, including the Raman inactive $B_{2g}^2$. This suggests that the Raman inactive LBMs ($B_{2g}^2$) in ENL-MoS$_2$ might be observed, with polarization behavior identical to the $A_1$$^{'}$ (LBMs) in ONL-MoS$_2$.

We now consider the evolution of the rigid-layer vibrations with increasing N based on symmetry analysis. In Figs\ref{fig:4}a,c, one C mode ($E_{2g}^2$) and one LBM ($B_{2g}^2$) are observed in 2L-MoS$_2$. Each splits in two branches with increasing N, one stiffening, the other softening with N. A new mode appears when N increases up to $4N+2$, $N=1,2,3,...$, and it splits into two branches again for higher N. The C modes ($E^{'}$) (Fig.\ref{fig:4}b) and LBMs ($A_1^{'}$) (Fig.\ref{fig:4}d) in ONL-MoS$_2$ exhibit similar trends with N as the C modes ($E_{2g}^2$)(Fig.\ref{fig:4}a) and LBMs ($B_{2g}^2$)(Fig.\ref{fig:4}c) in ENL-MoS$_2$, but with decreasing frequency for $E^{'}$ and increasing for $A_1^{'}$. Connecting each branch with solid lines shows that these form series of cone-like curves, Fig.\ref{fig:4}a-d. The number of LB and shear modes in ONL- and ENL-MoS$_2$ increases with N. However, in the experiment, no more than 3 of them are observed. In both ONL- and ENL-MoS$_2$, most of the observed shear modes are from the upper branch, and their frequencies stiffen with increasing N, while most of the LBMs are from the lower branch, and their frequencies soften with increasing N.
\begin{figure}[tb]
\centerline{\includegraphics[width=80mm,clip]{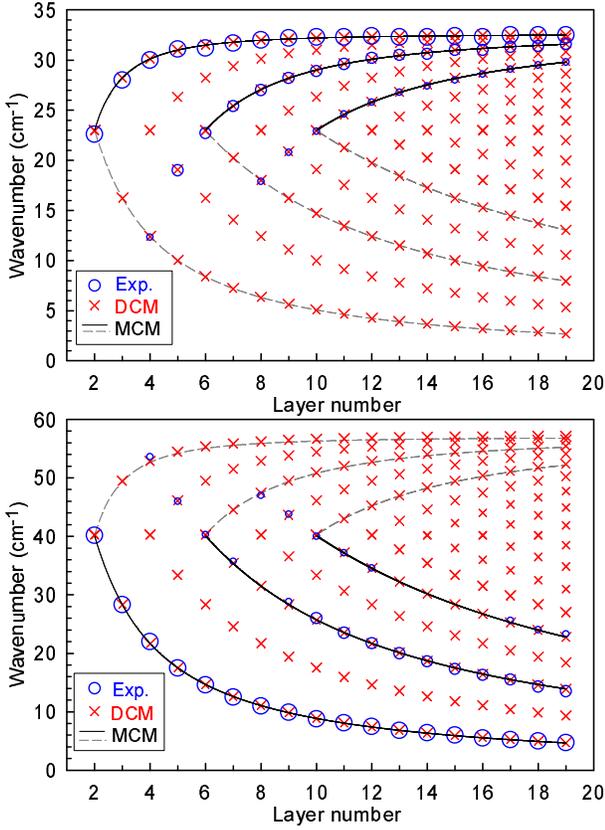}}
\caption{Position of (a) shear modes and (b) LBMs as a function N. The blue open circles are the experimental data. The diameter of the circles represents the Raman intensity of each mode. The red crosses are calculations based on the diatomic chain model. The black solid lines and gray dashed lines are, respectively, fitted by $\omega(N)=\omega(2N_0)\sqrt{1+cos(N_0\pi/{N})}$ (N$\ge$2N$_0$) and $\omega(N)=\omega(2N_0)\sqrt{1-cos(N_0\pi/{N})}$ (N$\ge$2N$_0$) for the branches originating from 2L-(N$_0$=1), 6L-(N$_0$=3) and 10L(N$_0$=5)-MoS$_2$ based on the monatomic chain model}
\label{fig:5}
\end{figure}

Fig.\ref{fig:5} plots the positions of all the observed and calculated rigid-layer vibration modes. The shear mode at 22.6cm$^{-1}$($E_{2g}^2$) in 2L-MoS$_2$ blue-shifts to 28cm$^{-1}$($E^{'}$) in 3L-MoS$_2$, and to 32.5cm$^{-1}$($E^{'}$) in 19L-MoS$_2$, reaching$\sim$32.7cm$^{-1}$ ($E_{2g}^2$) in bulk MoS$_2$. In multi-layer graphene\cite{tan12}, the ratio of C peak positions in bulk and 2LG is $\omega_{bulk}/\omega_{2LG}=\sqrt{2}$. In our MoS$_2$ measurements we have $32.7/22.6=1.447$, very close to $\sqrt{2}$. On the other hand, the LBM at 40cm$^{-1}$($B_{2g}^2$) in 2L red-shifts to 29cm$^{-1}$($A_1^{'}$) in 3L-MoS$_2$, and 5cm$^{-1}$($A_1^{'}$) in 19L-MoS$_2$. The blue-shifted branch reaches 56.8cm$^{-1}$ in 19L-MoS$_2$, close to the LBM value in bulk MoS$_2$\cite{Wakabayashi}. The relative displacements in 2L-MoS$_2$ between Mo and two S atoms within one plane for rigid-layer vibration modes are around 0.6$\%$, decreasing with increasing N. We can thus further simplify the model collapsing an entire layer in a single ball. So we consider a reduced monatomic chain model (MCM). Taking one layer as a ball with mass ($m_{Mo}+2m_s$) and interlayer bonding $\alpha_{ss}^\perp$ for LBMs, and $\alpha_{ss}^\parallel$ for shear modes, we get: $\omega_{LBM}=(1/\sqrt{2}\pi c)\sqrt{\alpha_{ss}^\perp/(m_{Mo}+2m_s)}$, $\omega_C=(1/\sqrt{2}\pi c)\sqrt{\alpha_{ss}^\parallel/(m_{Mo}+2m_s)}$ for 2L-MoS$_2$, with $\omega_{LBM}$ the LBM position, and $\omega_C$ the C peak position. The corresponding $\omega_{LBM}$ (40.8cm$^{-1}$) and $\omega_C$ (23.0cm$^{-1}$) are in good agreement with those from the DCM ($\omega_{LBM}$=40.3 cm$^{-1}$ and $\omega_C$=22.9cm$^{-1}$) and the experimental data ($\omega_{LBM}$=40.2 cm$^{-1}$ and $\omega_C$=22.6cm$^{-1}$). We can now solve the eigen-equation analytically and find the relation between position and N both for shear and LB modes. These vibration modes can be assigned to several branches, as shown in Fig.\ref{fig:5}. A new branch will emerge from each ENL-MoS$_2$, i.e. 2,4,6,8L...-MoS$_2$, at about the same position as that of the C mode or LBM in 2L-MoS$_2$, then splitting into two subbranches, one blue-shifting, the other red-shifting with increasing N. The observed C modes are usually in the high-frequency subbranch, while the corresponding LBMs are usually in the low-frequency one; these are connected by solid lines in Fig.\ref{fig:5}. For the branches originating from each ENL-MoS$_2$, the frequency as a function of N is $\omega(N)=\omega(2N_0)\sqrt{1+cos(N_0\pi/{N})}$ (N$\ge$2N$_0$, with N$_0$ an integer: 1,2,3,4....) for the high-frequency subbranch, and $\omega(N)=\omega(2N_0)\sqrt{1-cos(N_0\pi/{N})}$ (N$\ge$2N$_0$) for the low-frequency one. E.g., for the high-frequency C subbranch originating from 2L-MoS$_2$, we have N$_0$=1, thus $\omega_C(N)=\omega_C(2)\sqrt{1+cos(\pi/{N})}$ (N$\ge$2), where $\omega_C(2)$=23.0 cm$^{-1}$. If we replace $\omega_C(2)\sim$23.0cm$^{-1}$ in MoS$_2$ with $\omega_C(2)\sim$31cm$^{-1}$, this relation describes the C peaks in multilayer graphenes\cite{tan12}. Similarly, $\omega_C(N)=\omega_C(6)\sqrt{1+cos(3\pi/{N})}$ (N$\ge$6) for the high-frequency subbranch originating from 6L-MoS$_2$, and $\omega_C(N)=\omega_C(10)\sqrt{1+cos(5\pi/{N})}$ (N$\ge$10) for the high-frequency subbranch originating from 10L-MoS$_2$, where $\omega_C(6)$ and $\omega_C(10)$ are almost the same as $\omega_C(2)$. The observed LBMs in MoS$_2$ mainly come from the low-frequency subbranches. The relation between frequency and N in the low-frequency subbranch originating from 2, 6 and 10L-MoS$_2$ are: $\omega_{LBM}(N)=\omega_{LBM}(2)\sqrt{1-cos(\pi/{N})}$ (N$\ge$2), $\omega_{LBM}(N)=\omega_{LBM}(6)\sqrt{1-cos(3\pi/{N})}$ (N$\ge$6) and $\omega_{LBM}(N)=\omega_{LBM}(10)\sqrt{1-cos(5\pi/{N})}$ (N$\ge$10), where $\omega_{LBM}(2)$=40.8cm$^{-1}$. Also, $\omega_{LBM}(6)$ and $\omega_{LBM}(10)$ are almost the same as $\omega_{LBM}(2)$. These relations match well the experiments, Figs\ref{fig:5}a,b. Fig.\ref{fig:5} also shows another subbranch for 2, 6 and 10L-MoS$_2$ both for shear and LB modes indicated by gray dashed lines. Only one or two modes are detected for these subbranches, with positions in good agreement with the model predictions.

Note that any coupling between supported MoS$_2$ and the substrate is not included in our chain models. The excellent agreement between experiments and model predictions means that the coupling between MoS$_2$ and the substrate does not play a major role, the scaling with N being only determined by the interaction between the MoS$_2$ layers. Indeed, for the suspended multilayer graphene in Ref.[\onlinecite{tan12}], no coupling was considered, and the C scaling with N was also well described by a MCM.

In principle, our chain model can be extended to predict rigid-layer vibrations in other LMs. The general approach is to calculate the reduced mass for the monolayer of a given material, and then measure C and LBMs in 2L samples. One can then predict the relation between frequency and N for the different branches in any LM. E.g., the theoretical positions of the C and LB modes in 2L-hBN are$\sim$38.6cm$^{-1}$ and 85.6cm$^{-1}$, respectively\cite{Michel12}. Our model predicts that the C mode generates two branches, $\omega(N)=38.6\sqrt{1+cos(\pi/{N})}$ (N$\ge$2) at higher frequency, and $\omega(N)=38.6\sqrt{1-cos(\pi/{N})}$ (N$\ge$2) at lower. The LBM also generates two branches, $\omega(N)=85.6\sqrt{1+cos(\pi/{N})}$ (N$\ge$2) at higher-frequency, and $\omega(N)=85.6\sqrt{1-cos(\pi/{N})}$ (N$\ge$2) at lower. Similarly, the C mode ($\sim$38.6cm$^{-1}$) and LBM ($\sim$85.6cm$^{-1}$) in 4L-hBN, will also generate two branches, and so on. Thus, we can predict all the C and LB modes in NL-hBN.

In conclusion, we characterized single and few layer MoS$_2$ by Raman spectroscopy. We observed rigid-layer vibrations both for shear and layer breathing modes, assigned to the irreducible representations of the point group which the sample belongs to, as confirmed by polarized Raman spectroscopy. These change with number of layers, with different scaling for odd and even layers. A diatomic chain model, combined with group theory, explains the observed trends. Furthermore, a reduced monatomic chain model can be used to describe the shear and layer breathing modes in MoS$_2$ and any other layered material with any number of layers.

We acknowledge support from special funds for Major State Basic Research of China, No. 2009CB929301, the National Natural Science Foundation of China, grants 11225421 and 10934007, EU grants NANOPOTS and GENIUS, EPSRC grant EP/G042357/1, and a Royal Society Wolfson Research Merit Award.


\begin{thebibliography}{99}

\bibitem{novos2d} Novoselov, K. S.; Castro Neto, A. H. Phys. Scr. T \textbf{2012}, 146, 014006.

\bibitem{bonacm} Bonaccorso, F.; Lombardo, A., Hasan, T.; Sun, Z.; Colombo, L.; Ferrari, A. C.  Materials Today \textbf{15}, 14 (2012)

\bibitem{coleman} Coleman, J. N., et al. Science \textbf{2011}, 331, 568.

\bibitem{Wilson1969} Wilson, J. A.; Yoffe, A. D.  Adv. Phys. \textbf{1969},18, 193.

\bibitem{Poizot2000} Poizot, P.; Laruelle, S.; Grugeon, S.; Dupont, L.; Tarascon, J. M.  Nature \textbf{2000}, 407, 496-499.

\bibitem{Prasad1997} Prasad, S. V.; Zabinski, J. S. Nature \textbf{1997}, 387, 761-763.

\bibitem{Abruna1982} Abruna, H. D.; Bard, A. J. J. Electrochem. Soc. \textbf{1982}, 129, 673.

\bibitem{frindt_jap_1966} Frindt, R. F.  J. App. Phys. \textbf{1966}, 37, 1928.

\bibitem{Clement1978} Clement, R. P.  Inorg. Chem. \textbf{1978}, 17, 2754.

\bibitem{Frey2003} Frey, G. L.; Reynolds, K. J.; Friend, R. H.; Cohen, H.; Feldman, Y.  J. Am. Chem. Soc. \textbf{2003}, 125, 5998-6007.

\bibitem{Mishra1997} Mishrayz, S. K.; Satpathyyz, S.; Jepsenz, O.  J. Phys. Condens. Matter \textbf{1997}, 9, 461-470.

\bibitem{Kane2005} Kane, C. L.; Mele, E. J.  Phys. Rev. Lett. \textbf{2005}, 95, 226801.

\bibitem{kis_nn_2011} Radisavljevic, B.; Radenovic, A.; Brivio, J.; Giacometti, V.; Kis, A.  Nat. Nanotech. \textbf{2011}, 6, 147.

\bibitem{heinz_prl_2010} Mak, K. F.; Lee, C.; Hone, J.; Shan, J.; Heinz, T. F.  Phys. Rev. Lett. \textbf{2010}, 105, 136805.

\bibitem{fenwang_nl_2010} Splendiani, A.; Sun, L.; Zhang, Y.; Li, T.; Kim, J.; Chim, C. Y.; Galli, G.; Wang, F.  Nano Lett. \textbf{2010},10, 1271.

\bibitem{sundaram} Sundaram, R. S.; Engel, M.; Lombardo, A.; Krupke, R.; Ferrari, A. C.; Avouris, P.; Steiner, M.  arXiv:1211.4311.

\bibitem{heinz_nn_2012} Mak, K. F.; He, K.; Shan, J.; Heinz, T. F.  Nat. Nanotech. \textbf{2012}, 7, 494-498.

\bibitem{Cui12} Zeng, H.; Dai, J.; Yao, W.; Xiao, D.; Cui, X. Nat. Nanotech. \textbf{2012}, 7, 490-493.

\bibitem{Feng12} Cao, T.; Wang, G.; Han, W. P.; Ye, H. Q.; Zhu, C. R.; Shi, J. R.; Niu, Q.; Tan, P. H.; Wang, E.; Liu, B. L.; Feng, J. Nat. Commun. \textbf{2012}, 3, 887.

\bibitem{pollmann_prb_2001} B\"{o}ker, T.; Severin, R.; M\"{u}ller, A.; Janowitz, C.; Manzke, R.; Vo{\ss}, D.; Kr\"{u}ger, P.; Mazur, A.; Pollmann, J.  Phys. Rev. B \textbf{2001}, 64, 235305.

\bibitem{wold_prb1987} Coehoorn, R.; Haas, C.; Dijkstra, J.; Flipse, C. J. F.; de Groot, R. A.; Wold, A. Phys. Rev. B \textbf{1987}, 35, 6195.

\bibitem{groot_prb_1987} Coehoorn, R.; Haas, C.; de Groot, R. A. Phys. Rev. B  \textbf{1987}, 35, 6203.

\bibitem{mattheiss_prb_1973} Mattheiss, L. F. Phys. Rev. B \textbf{1973}, 8, 3719.

\bibitem{lieber_apl_1991} Kim, Y.; Huang, J.-L.; Lieber, C. M.  Appl. Phys. Lett. \textbf{1991}, 59, 3404.

\bibitem{novoselov_s_2004} Novoselov, K. S. Science \textbf{2004}, 306, 666.

\bibitem{parkison_jpc_1092} Kam, K. K.; Parkinson, B. A.  J. Phys. Chem. \textbf{1982}, 86, 463-467.

\bibitem{galli_jpcc_2007} Li, T.; Galli, G.;  J. Phys. Chem. C \textbf{2007}, 111, 16192.

\bibitem{iwasa_nl_2012} Zhang, Y.; Ye, J.; Matsuhashi, Y.; Iwasa, Y. Nano Lett. \textbf{2012}, 12, 1136-1140.

\bibitem{pdye_IEEE_2012} Liu, H.; Ye, P. D. IEEE Electron Dev. Lett. \textbf{2012}, 33, 546.

\bibitem{pdye_acs_2012} Liu, H.; Neal, A. T.; Ye, P. D.  ACS Nano \textbf{2012}, 6, 8563-8569.

\bibitem{zhang_acs_2011} Yin, Z.; Li, H.; Li, H.; Jiang, L.; Shi, Y.; Sun, Y.; Lu, G.; Zhang, Q.; Chen, X.; Zhang, H.  ACS Nano \textbf{2011}, 6, 74-80.

\bibitem{orta_oe_2012} Alkis, S.; \"{O}zta, T.; Ayg\"{u}n, L. E.; Bozkurt, F.; Okyay, A. K.; Orta\c{c}, B. Opt. Express. \textbf{2012}, 20, 21815.

\bibitem{im_nl_2012} Lee, H. S.; Min, S. W.; Chang, Y. G.; Park, M. K.; Nam, T.; Kim, H.; Kim, J. H.; Ryu, S.; Im, S.  Nano Lett. \textbf{2012}, 12, 3695-3700.

\bibitem{ferraribook} Raman spectroscopy in carbons: From nanotubes to diamond, Ferrari, A. C.; Robertson, J.  Philos. Trans. R. Soc. Lond. A, 362, 2477-2512, 2004.

\bibitem{ferrari07} Ferrari, A. C. Solid State Communications \textbf{2007}, 143, 47-57.

\bibitem{Tuinstra1970} Tuinstra, F.; Koenig, J. L.  J. Chem. Phys. \textbf{1979} 53, 1126.

\bibitem{Nemanich1979} Nemanich, R. J.; Solin, S. A. Phys. Rev. B \textbf{1979} 20, 392.

\bibitem{wangyan1990} Yan, W.; Alsmeyer, D. C.; McCeery, R. L. Chem. Mater. \textbf{1990} 2, 557.

\bibitem{ferrari06} Ferrari, A. C.; Meyer, J. C.; Scardaci, V.; Casiraghi, C.; Lazzeri, M.; Mauri, F.; Piscanec, S.; Jiang, D.; Novoselov, K. S.; Roth, S.; Geim, A. K. Phys. Rev. Lett. \textbf{2006}, 97, 187401.

\bibitem{tan12} Tan, P. H.; Han, W. P.; Zhao, W. J.; Wu, Z. H.; Kai, C.; Wang, H.; Wang, Y. F.; Bonini, N.; Marzari, N.; Savini, G.; Lombardo, A.; Ferrari, A. C. Nat. Mater. \textbf{2012}, 11, 294.

\bibitem{tony nl 2012} Lui, C. H.; Malard, L. M.; Kim, S.; Lantz, G.; Laverge, F. E.; Saito, R.; Heinz, T. F. Nano Lett. \textbf{2012}, 12, 5539-5544.

\bibitem{tony nl 2012b} Lui, C. H.; Heinz. T. F. arXiv:1210.0960.

\bibitem{sato2011} Sato, K.; Park, J. S.; Saito, R.; Cong, C.; Yu, T.; Lui, C. H.; Heinz, T. F.; Dresselhaus, G.; Dresselhaus, M. S. Phys. Rev. B \textbf{2011}, 84, 035419.

\bibitem{Latil} Latil, S.; Meunier, V.; Henrard, L. Phys. Rev. B \textbf{2007}, 76, 201402(R).

\bibitem{McCann} McCann, E. Phys. Rev. B \textbf{2006}, 74, 161403(R).

\bibitem{Koshino} Koshino, M.; Ando, T. Solid State Commun. \textbf{2009}, 149, 1123-1127.

\bibitem{sekine1980} Sekine, T.; Nakashizu, T.; Toyoda, K.; Uchinokura, K.; Matsuura, E. Solid State Comm. \textbf{1980}, 35, 371-373.

\bibitem{reich05} Reich, S.; Ferrari, A. C. Phys. Rev. B \textbf{2005}, 71, 205201.

\bibitem{Heinz10b} Lee, C.; Yan, H.; Brus, L. E.; Heinz, T. F.; Hone, J.; Ryu, S. ACS Nano \textbf{2010}, 4, 2695-2700.

\bibitem{jzhang2011} Zhang, J.; Peng, Z.; Soni, A.; Zhao, Y.; Xiong, Y.; Peng, B.; Wang, J.; Dresselhaus, M. S.; Xiong, Q. Nano Lett. \textbf{2011}, 11, 2407-2414.

\bibitem{Plechinger2012}Plechinger, G.; Heydrich, S.; Eroms, J.; Weiss, D.; Schueller, C.; Korn, T. Appl. Phys. Lett. 2012, 101, 101906.

\bibitem{cuixd12} Zeng, H.; Zhu, B.; Liu, K.; Fan, J.; Cui, X.; Zhang, Q. M. Phys. Rev. B \textbf{2012}, 86, 241301(R).

\bibitem{wiet70} Verble, J. L.; Wieting, T. J. Phys. Rev. B \textbf{1970}, 25, 362.

\bibitem{wiet71} Wieting, T. J.; Verble, J. L. Phys. Rev. B \textbf{1971}, 3, 4286.

\bibitem{Li12} Li, H.; Zhang, Q.; Ray Yap, C. C.; Tay, B. K.; Edwin, T. H. T.; Olivier, A.; Baillargeat, D. Adv. Funct. Mater. \textbf{2012}, 22, 1385-1390.

\bibitem{c44} Kuzuba, T.; Ishii, M. Phys. Stat. Sol. (b) \textbf{1989}, 155, K13.

\bibitem{Michel12} Michel, K. H.; Verberck, B. Phys. Rev. B \textbf{2012}, 85, 094303.

\bibitem{pointdif} Molina-S$\acute{\rm{a}}$nchez, A.; Wirtz, L. Phys. Rev. B \textbf{2011}, 84, 155413.

\bibitem{richard74} Zallen, R.; Slade, M. Phys. Rev. B \textbf{1974}, 9, 1627.

\bibitem{Luo96}Luo, N. S.; Ruggerone, P.; Toennies, J. P. Phys. Rev. B \textbf{1996}, 54, 5051.

\bibitem{Novoselov2005} Novoselov, K. S.; Geim, A. K.; Morozov, S. V.; Jiang, D.; Katsnelson, M. I.; Grigorieva, I. V.; Dubonos, S. V.; Firsov, A. A.  Nature \textbf{2005}, 438, 197-200.

\bibitem{Casiraghi07} Casiraghi, C.; Hartschuh, A.; Lidorikis, E.; Qian, H.; Harutyunyan, H.; Gokus, T.; Novoselov, K. S.; Ferrari, A. C. Nano Lett. \textbf{2007}, 7, 2711-2717.

\bibitem{coordination} Sandoval, S. J.; Yang, D.; Frindt, R. F.; Irwin, J. C. Phys. Rev. B \textbf{1991}, 44, 3955.

\bibitem{zhangguangyin1991} Lattice Vibration Spectroscopy, Zhang, G. Y.; Lan, G. X.; Wang, Y. F. Second ed.; High Education Press: China, 1991.

\bibitem{ataca11} Ataca, C.; Topsakal, M.; Akt$\ddot{\rm{u}}$rk, E.; Ciraci, S. J. Phys. Chem. C \textbf{2011}, 115, 16354.

\bibitem{Davydov} Ghosh, P. N.; Maiti, C. R. Phys. Rev. B \textbf{1983}, 28, 2237.

\bibitem{windom} Windom, B. C.; Sawyer, W. G.; Hahn, D. W. Tribology Letters \textbf{2011}, 3, 301-310.

\bibitem{tensor} Loudon, R. Adv. Phys. \textbf{2001}, 50, 813.

\bibitem{polar11} Dumcenco, O. D.; Su, Y. C.; Wang, Y. P.; Chen, K. Y.; Huang, Y. S.; Ho, C. H.; Tiong, K. K. C. J. Phys. \textbf{2011}, 49, 270.

\bibitem{kuok2000} Kuok, M. H.; Ng, S. C.; Rang, Z. L. Phys. Rev. B \textbf{2000}, 62, 12902.

\bibitem{c11} McSkimin, H. J.; Andreatch, P.  J. Appl. Phys. \textbf{1964}, 35, 2161.

\bibitem{Feldman81} Feldman, J. L. J. Phys. Chem. Solids \textbf{1981}, 42, 1029.

\bibitem{Wakabayashi} Wakabayashi, N.; Smith, H. G.; Nicklow, R. M. Bull. Am. Phys. Soc. \textbf{1972}, 17, 292.

\end{thebibliography}
\end{document}